\documentclass[a4paper]{jpconf}
\usepackage{graphicx}

\usepackage{citesort}

\begin{document}
\title{Measurement of neutron yield by 62 MeV proton beam on a thick Beryllium target}

\author{
R.~Alba$^1$,%
M.~Barbagallo$^2$,%
P.~Boccaccio$^3$,%
A.~Celentano$^4$,\\%
N.~Colonna$^2$,%
G.~Cosentino$^1$,%
A.~Del~Zoppo$^1$,%
A.~Di~Pietro$^1$,\\%
J.~Esposito$^3$,%
P.~Figuera$^1$,%
P.~Finocchiaro$^1$,%
A. Kostyukov$^5$,\\%
C.~Maiolino$^1$,%
M.~Osipenko$^4$,%
G.~Ricco$^4$,%
M.~Ripani$^4$,%
C.M.~Viberti$^4$,\\%
D.~Santonocito$^1$,%
M.~Schillaci$^1$%
}

\address{$^1$ Istituto Nazionale di Fisica Nucleare, Laboratori Nazionali del Sud, Catania 95123, Italy}
\address{$^2$ Istituto Nazionale di Fisica Nucleare, Sezione di Bari, Bari 70126, Italy}
\address{$^3$ Istituto Nazionale di Fisica Nucleare, Laboratori Nazionali di Legnaro, Legnaro 35020, Italy}
\address{$^4$ Istituto Nazionale di Fisica Nucleare, Sezione di Genova, Genoa 16146, Italy}
\address{$^5$ Moscow State University, Moscow 119992, Russia}

\ead{osipenko@ge.infn.it}

\begin{abstract}
In the framework of research on IVth generation reactors and high intensity neutron sources a low-power prototype neutron amplifier was recently proposed by INFN. It is based on a low-energy, high current proton cyclotron, whose beam, impinging on a thick Beryllium converter, produces a fast neutron spectrum. The world database on the neutron yield from thick Beryllium target in the 70 MeV proton energy domain is rather scarce. The new measurement was performed at LNS, covering a wide angular range from 0 to 150 degrees and an almost complete neutron energy interval. In this contribution the preliminary data are discussed together with the proposed ADS facility.
\end{abstract}

\section{Introduction}\label{sec:intro}
Nuclear fission is one of the main sources of the energy production. Limited availability of fissile materials and difficulties in nuclear waste disposal call for the research on fast neutron reactors. A promising solution is the Accelerator Driven System (ADS)~\cite{ADS}, a fast sub-critical reactor fed by an external neutron source. Sufficiently high neutron fluxes can be generated by a particle accelerator or by a fusion reactor.

The high power proton cyclotron being installed in Laboratori Nazionali di Legnaro of INFN for SPES project~\cite{SPES} offers a possibility to build a low-power ADS prototype for research purposes. The design of such prototype has been developed in the framework of INFN-E project~\cite{infn_e_ads}. It is based on 0.5 mA, 70 MeV proton beam
impinging on a thick Beryllium target and fast sub-critical core consisting of solid lead matrix and 60 $UO_2$ fuel elements, enriched to 20\% with $^{235}$U. Both, production target and reactor core, are cooled by continous helium gas flow. The ADS is expected to have the effective neutron multiplication factor $k_{eff}=0.946$, neutron flux $\phi=3\div 6\times 10^{12}$ n/cm$^2$/s and thermal power of 130 kW at 200$^\circ$ C.
Such facility would allow to study the kinetics and dynamics of the fast reactor core,
the burn-up and transmutation of radioactive wastes,
issues related to system safety and licensing.

The choice of the production target material is determined by the low beam energy of the cyclotron. $^9$Be bombarded by protons, in fact, provides an abundant neutron source whose spectrum is sufficiently hard to burn minor actinides. The design of the proposed ADS requires careful knowledge of neutron yield produced by the 70 MeV proton beam on $^9$Be target. The existing data in the given energy range are rather scarce and not detailed. The integrated yield at 70 MeV was measured in Ref.~\cite{Tilquin05}, while differential yields for few angles were measured at various beam energies in Refs.~\cite{Waterman79,Johnsen76,Almos77,Heintz77,Meier88,Madey77,Harrison80}.
This lack of data demanded a dedicated measurement of neutron yield produced by the proton beam on a thick $^9$Be target. The measurement was performed at Laboratori Nazionali del Sud (LNS) of INFN using the existing superconducting cyclotron~\cite{lnsrep10}.
The details of this experiment are discussed below together with some preliminary data.

\section{Experimental setup}\label{sec:setup}
\begin{figure}[h]
\includegraphics[bb=100 10 560 520, width=24pc, angle=270]{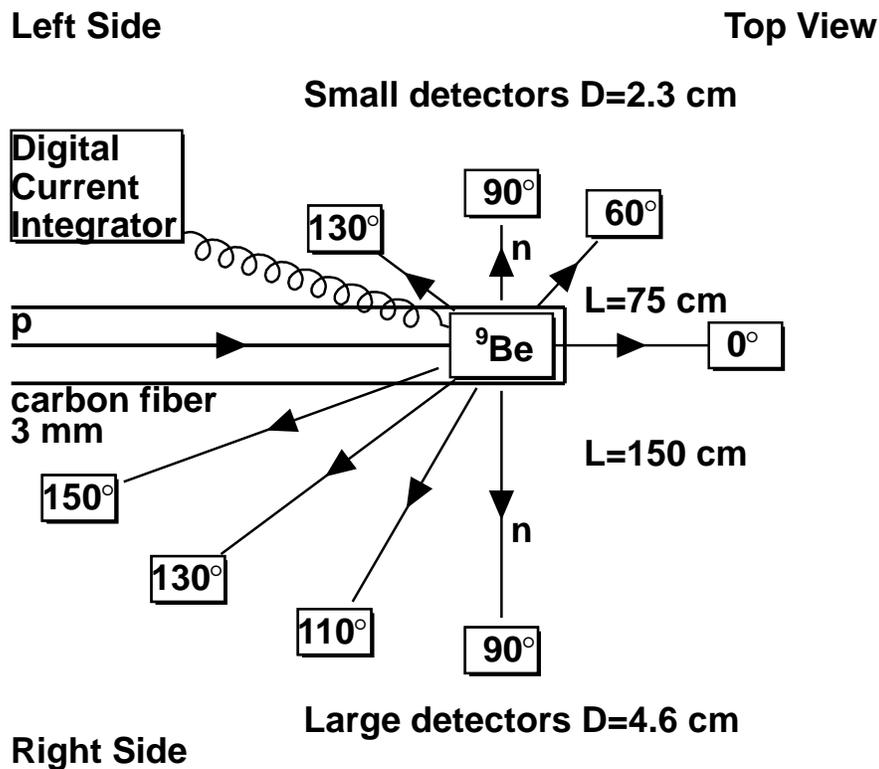}
\caption{\label{fig:setup}Scheme of experimental setup.}
\end{figure}

Proton beam of 62 MeV from LNS superconducting cyclotron was delivered in MEDEA experimental hall
with a beam current of 30-50 pA. The cyclotron beam structure (RF=40 MHz) was modified by suppression of
four bunches out of five. In this way 1.5 ns wide beam bunches arrived on the target
with period of 125 ns.
The target consisted of a solid 3 cm thick $^9$Be cylinder with 3.5 cm diameter. The target thickness
was chosen to ensure complete absorption of 70 MeV proton beam.
The electric charge deposited by the beam on the target was measured by a digital current integrator
and used for absolute normalization of the data.
The neutrons produced in the target were measure by Time-of-Flight (ToF) technique. To this end RF signal from
cyclotron was used as the reference time. Eight neutron detectors were installed simultaneously
at different angles and different distances around the target as shown for example in Fig.~\ref{fig:setup}.

Near detectors, installed at half distance from the target with respect to far detectors, were used to measure
low energy neutrons.

\subsection{Neutron detectors} \label{sec:det}
\begin{figure}[h]
\begin{minipage}{18pc}
\includegraphics[bb=10 360 475 800, width=18pc]{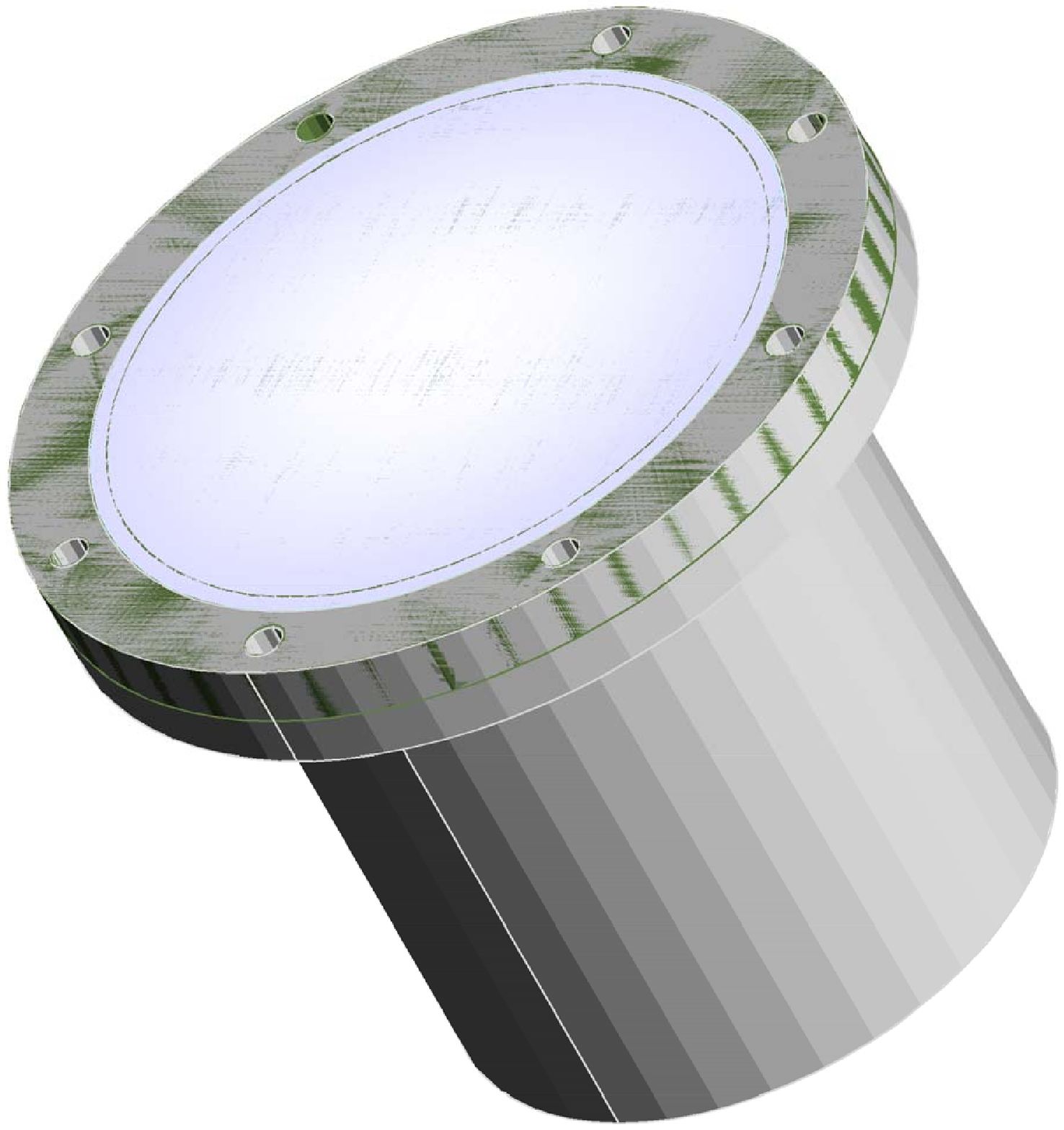}
\caption{\label{fig:det_l}Neutron detector drawing.}
\end{minipage}\hspace{2pc}%
\begin{minipage}{18pc}
\includegraphics[bb=20 170 580 640, width=18pc]{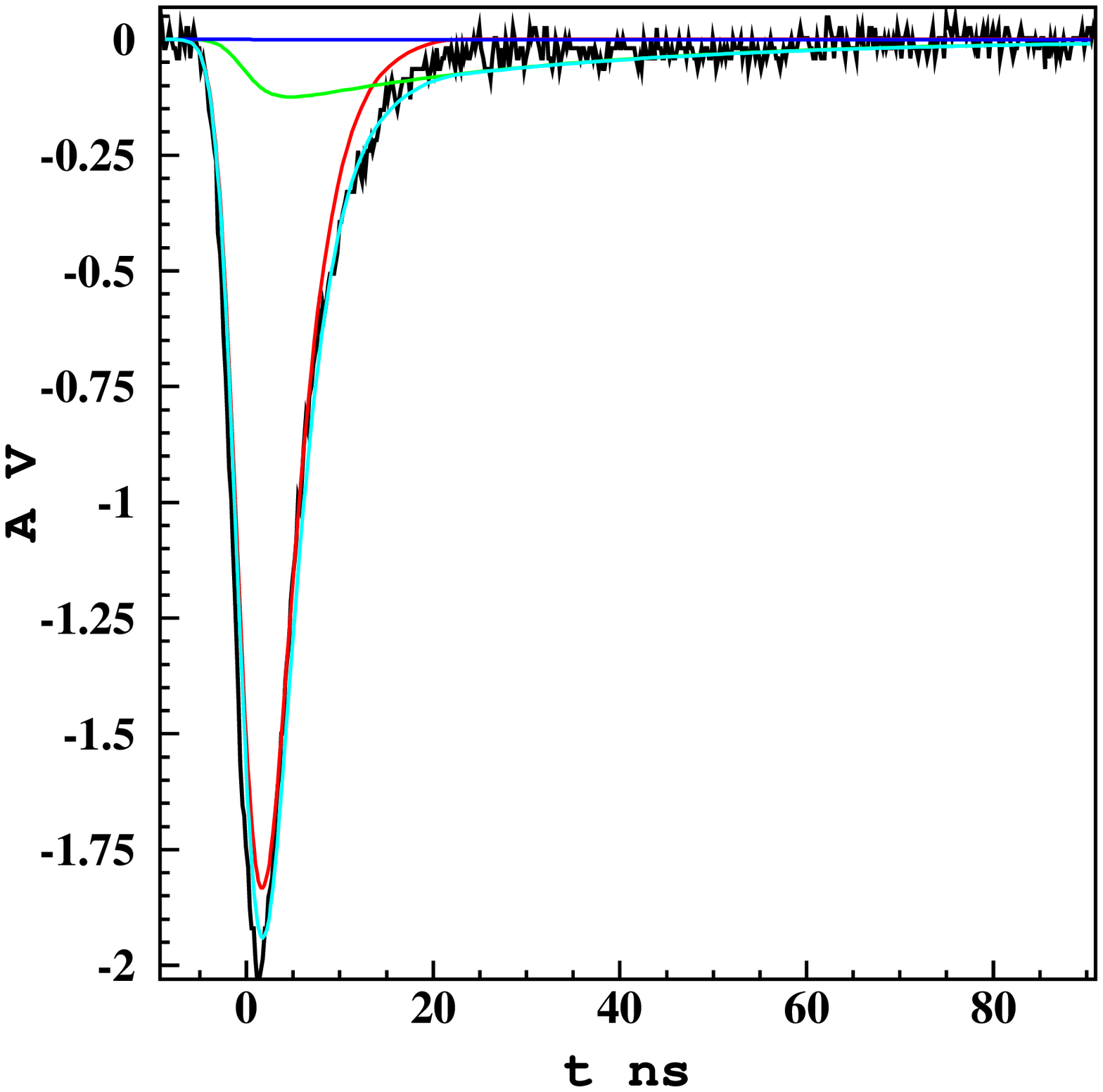}
\caption{\label{fig:det_r}Neutron detector response to $^{137}$Cs $\gamma$-source.
Red, green and blue curves show three main decay components of the scintillator,
while cyan curve represents their sum.}
\end{minipage}
\end{figure}

Neutron detectors were made of 4 cm long Aluminum cylindrical cells of two different diameters 4.6 cm and 2.3 cm
filled with EJ301 liquid scintillator. Each cell was sealed in Ar/N$_2$ atmosphere by a borosilicate glass.
Scintillation light was detected by ET9954B PhotoMultiplier Tube (PMT) coupled to the cell glass
by optical grease. PMT voltage divider with a large linearity range was selected to allow measurements of signals
spread by few orders of magnitude.
CAD drawing of the cell is shown in Fig.~\ref{fig:det_l}.
EJ301 liquid scintillator produces fast rise-time signals, as shown for example in Fig.~\ref{fig:det_r}.
This feature is necessary for a good ToF resolution. Furthermore the scintillator has a good Pulse Shape
Discrimination capability, necessary for $\gamma$/n separation.

\subsection{Data acquisition system}\label{sec:daq}
Two independent data acquisition systems were used for two types of detectors.
These systems were based on CAMAC LeCroy FERA modules connected to Linux PC
via GPIB interface. The system dead-time did not exceed 13$\mu$s allowing
for high data rates of few kHz on each detector to be measured.

Two different types of detectors had different dynamic ranges:
near detectors were measuring signals with deposited energy $<$0.5 MeVee,
while far detectors were configured to cover complete deposited energy range $<$50 MeVee.
Thus near detectors had an expanded low-deposited energy domain, reducing
the detection threshold and improving $\gamma$/n-separation for small signals.

\section{Data analysis}\label{sec:ana}
The measured data were analyzed off-line to combine different observables,
remove identified backgrounds, correct the data for detector inefficiencies.
The efficiencies of neutron detectors were obtained by means of Geant4
Monte Carlo simulations. Experimental verification of these simulations
with a well known neutron source are foreseen in near future.
Some details of the data analysis are described below.

\subsection{Detector calibrations}\label{sec:calib}
Neutron detectors were calibrated using a number of $\gamma$-sources through the measurement
of the backward Compton scattering. To this end the coincidences of recoil electron in liquid
scintillator with scattered $\gamma$ in BaF$_2$ crystal were measured.
To obtain a good precision we employed the following variety of $\gamma$-sources:
$^{207}$Bi, $^{137}$Cs, $^{60}$Co, $^{22}$Na, $^{152}$Eu and $^{241}$Am.
\begin{figure}[h]
\begin{minipage}{18pc}
\includegraphics[bb=20 180 580 640, width=18pc]{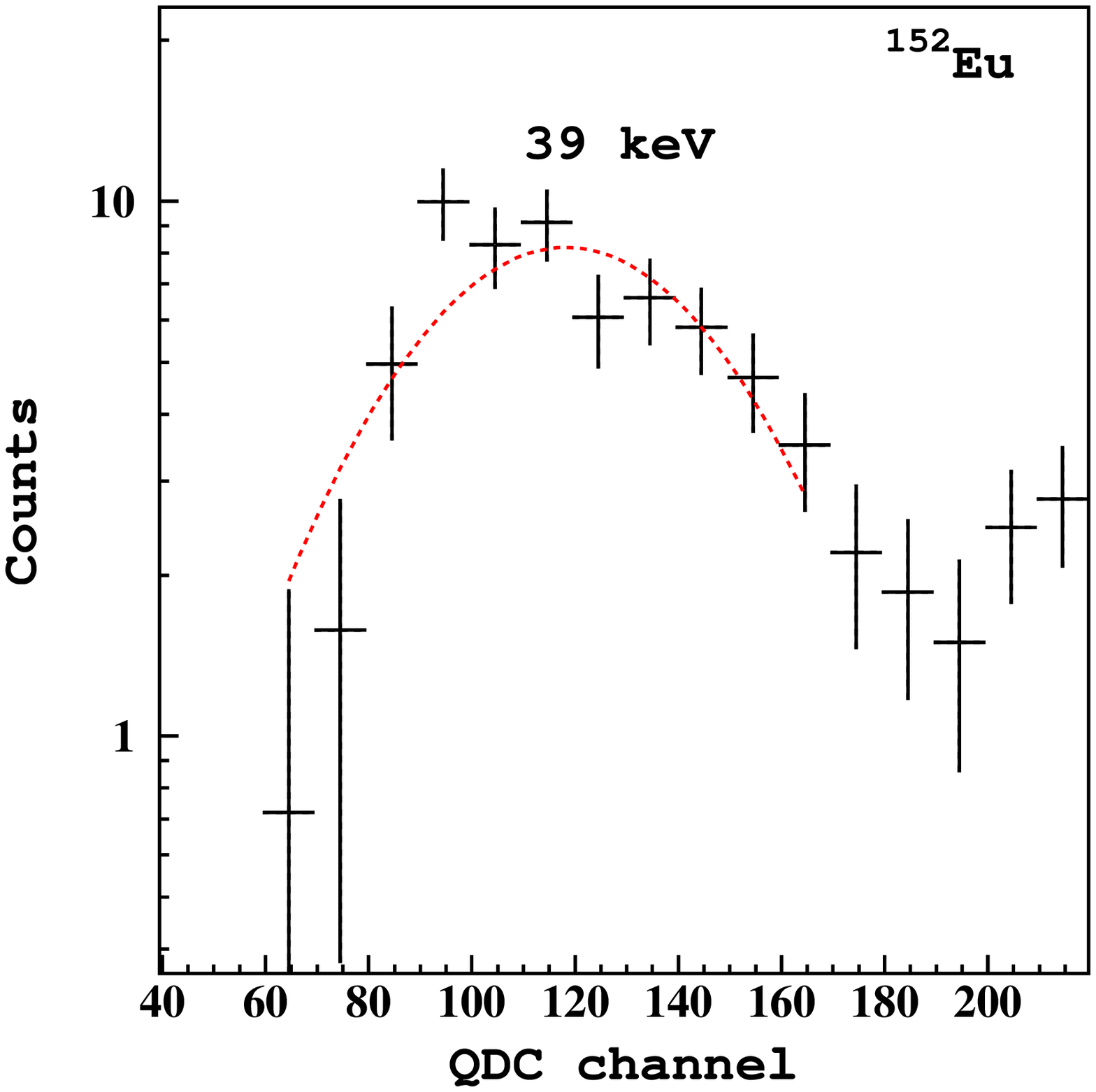}
\caption{\label{fig:calib_small_l}Backward Compton scattering spectrum with $^{152}$Eu $\gamma$-source in near detector.
Dashed red curve shows Gaussian fit.}
\end{minipage}\hspace{2pc}%
\begin{minipage}{18pc}
\includegraphics[bb=20 180 580 640, width=18pc]{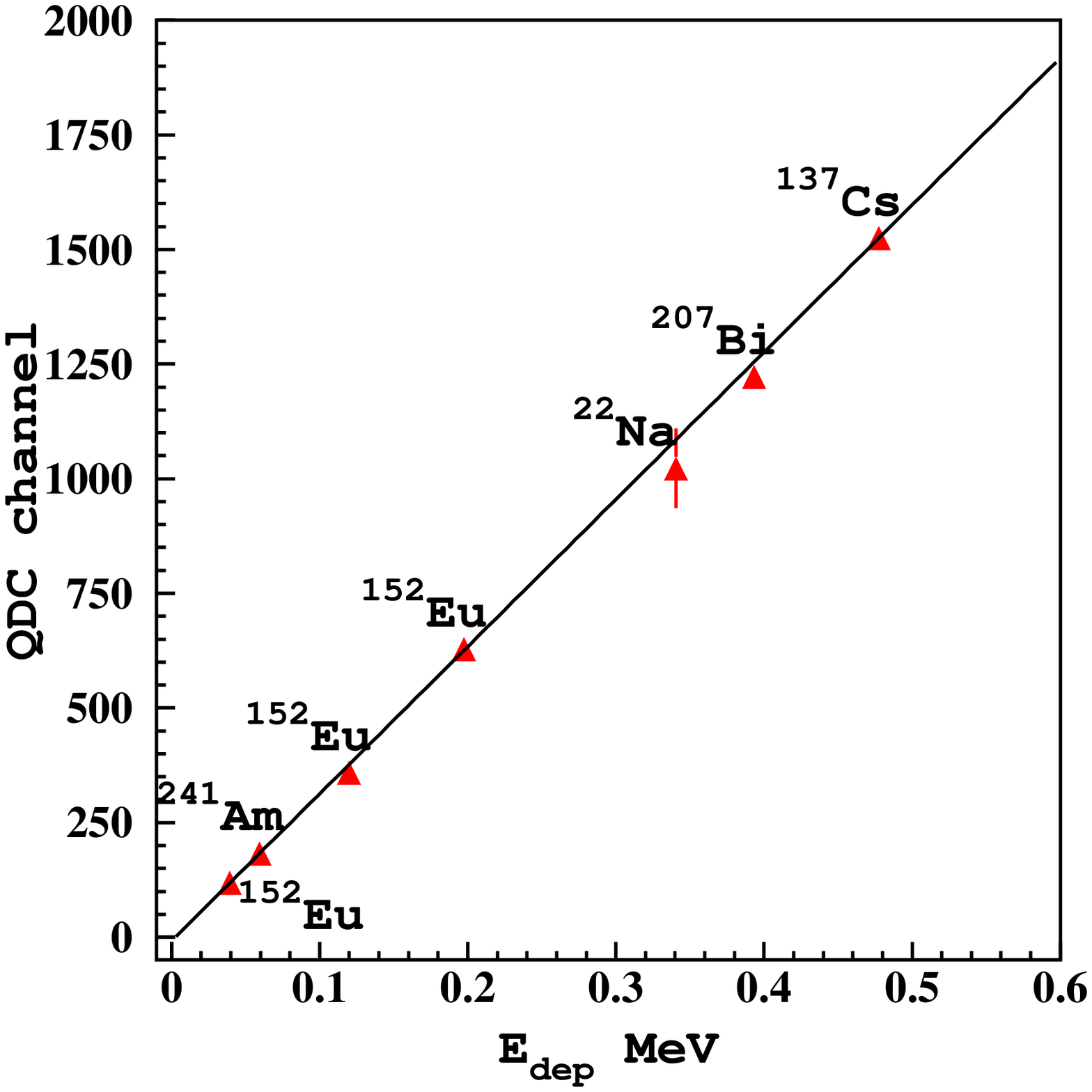}
\caption{\label{fig:calib_small_r}Calibration of near neutron detector with $\gamma$-sources. Black line shows linear fit to the data.}
\end{minipage}
\end{figure}
\vspace{-1.5pc}%
\begin{figure}[h]
\begin{minipage}{18pc}
\includegraphics[bb=20 180 580 640, width=18pc]{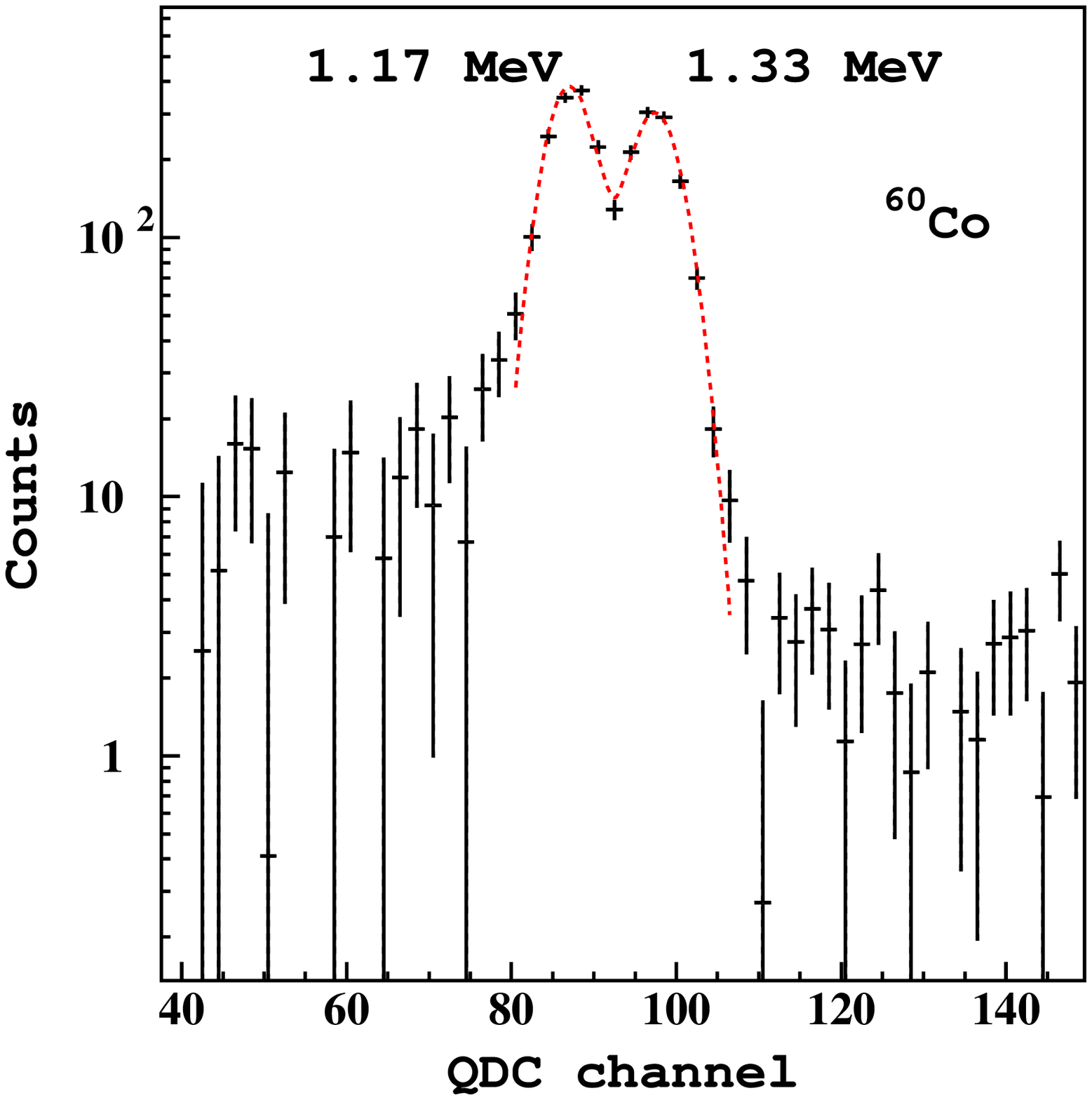}
\caption{\label{fig:calib_large_l}Backward Compton scattering spectrum with $^{60}$Co $\gamma$-source in far detector.
Dashed red curve shows double Gaussian fit.}
\end{minipage}\hspace{2pc}%
\begin{minipage}{18pc}
\includegraphics[bb=20 180 580 640, width=18pc]{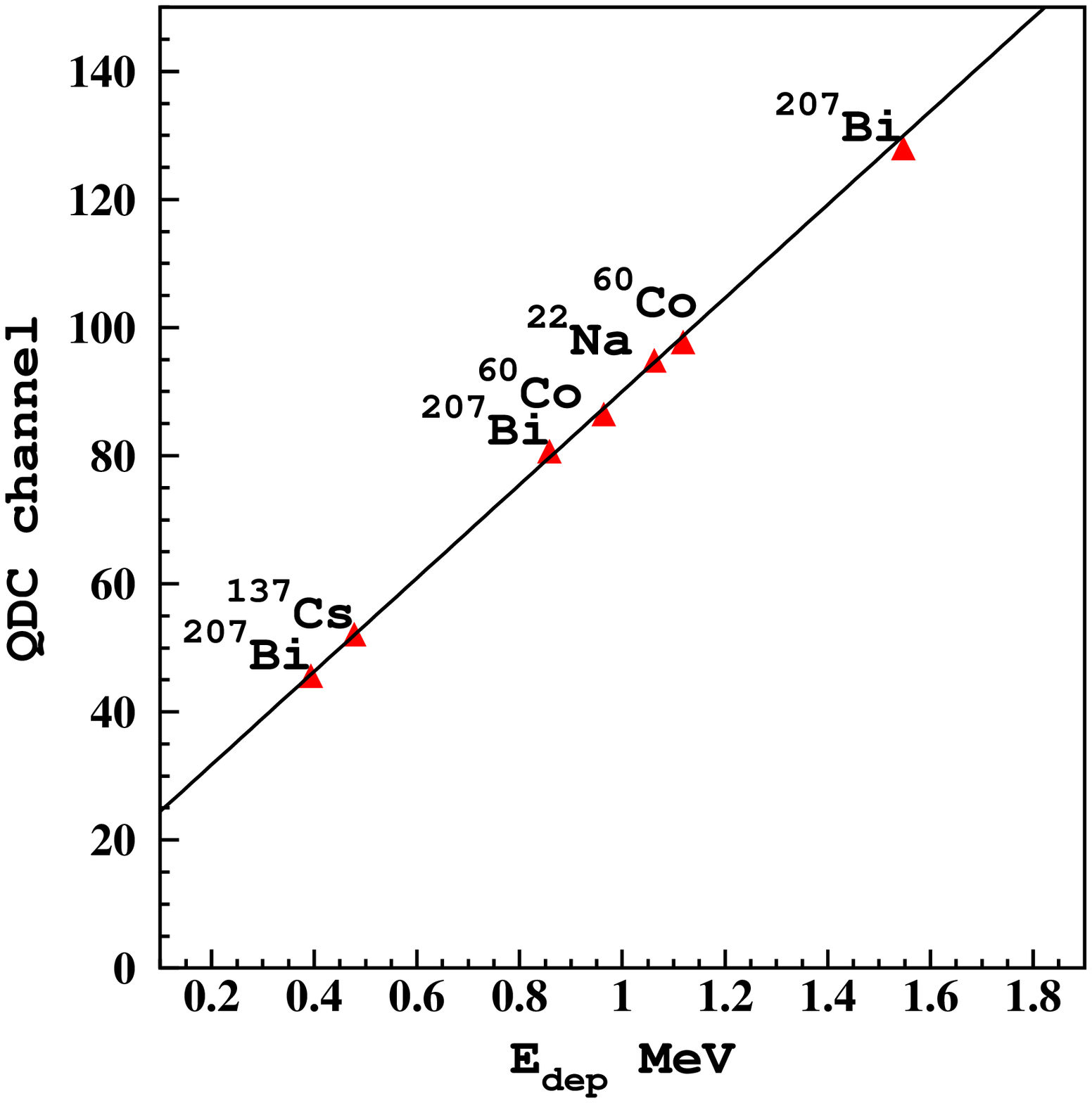}
\caption{\label{fig:calib_large_r}Calibration of far neutron detector with $\gamma$-sources. Black line shows linear fit to the data.}
\end{minipage}
\end{figure}
Examples of backward Compton scattering peak fitting and calibration are shown in Figs.~\ref{fig:calib_small_l}
and \ref{fig:calib_small_r} for near detectors and in Figs.~\ref{fig:calib_large_l}
and \ref{fig:calib_large_r} for far detectors.
The neutron detectors exhibited a good deposited energy resolution $\Delta E_{dep}/E_{dep}<$5\%/$\sqrt{E_{dep}}$.

\subsection{Background removal}\label{sec:bkg}
The measured event yields contained two main types of background:
\begin{enumerate}
\item environmental background triggered by particles coming not from the target,
but from surrounding materials as for example the beam pipe, detector and beam support structures
and hall floor;
\item contamination of $\gamma$s produced by nuclear reactions after some delay. The prompt
$\gamma$s do not contaminate measured spectrum because they show up as a separate peak in the ToF spectrum.
\end{enumerate}
\begin{figure}[h]
\includegraphics[bb=20 180 580 640, width=18pc]{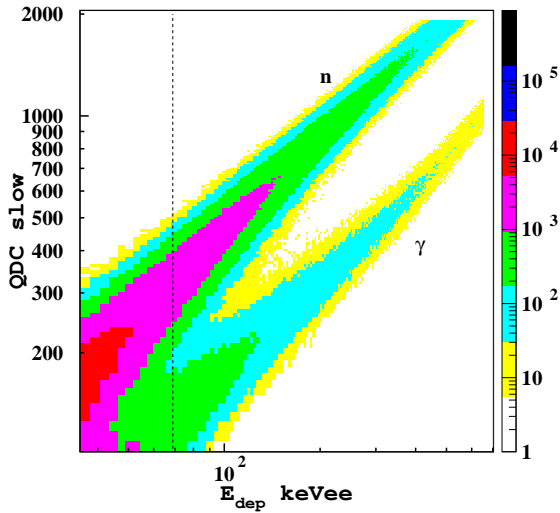}\hspace{2pc}%
\begin{minipage}[b]{18pc}\caption{\label{fig:psd}Pulse shape discrimination of neutrons.}
\end{minipage}
\end{figure}
The first type of background had been measured in special runs by obscuring the target solid angle,
as seen from the detector, by a shadow bar. The data from these special runs were analyzed in the
same way as production run data and the obtained yields were subtracted.
The second background was disentangled by PSD technique. To this end the integral of the signal pulse tail
(QDC slow) was recorded for each event. PSD was performed by a comparison of the signal pulse tail contribution
to the total visible deposited energy.
An example of PSD in near detector is shown in Fig.~\ref{fig:psd}.
\begin{figure}[h]
\begin{minipage}{18pc}
\includegraphics[bb=20 180 580 640, width=18pc]{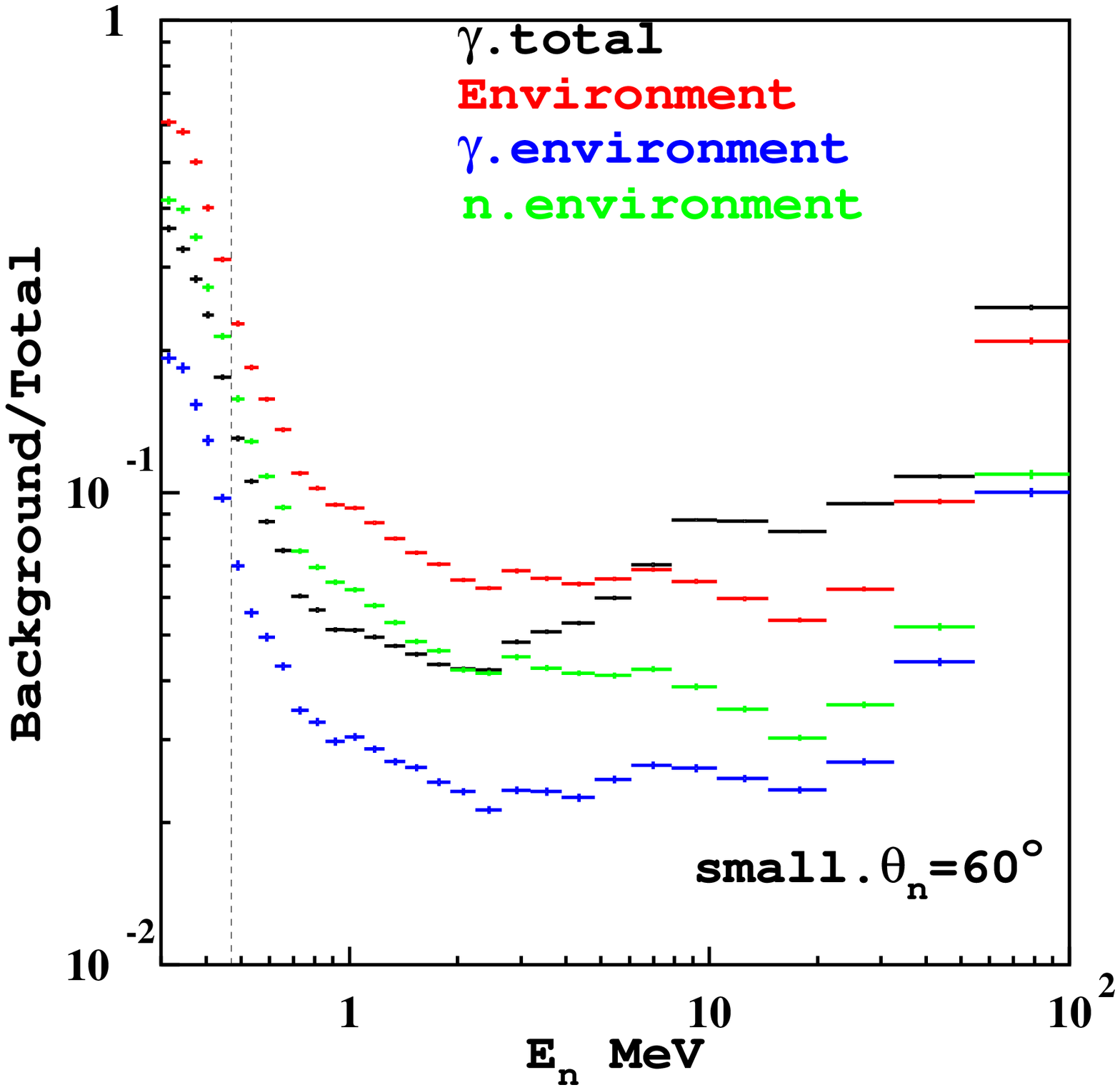}
\caption{\label{fig:bkg_l}Background contributions for near detectors.}
\end{minipage}\hspace{2pc}%
\begin{minipage}{18pc}
\includegraphics[bb=20 180 580 640, width=18pc]{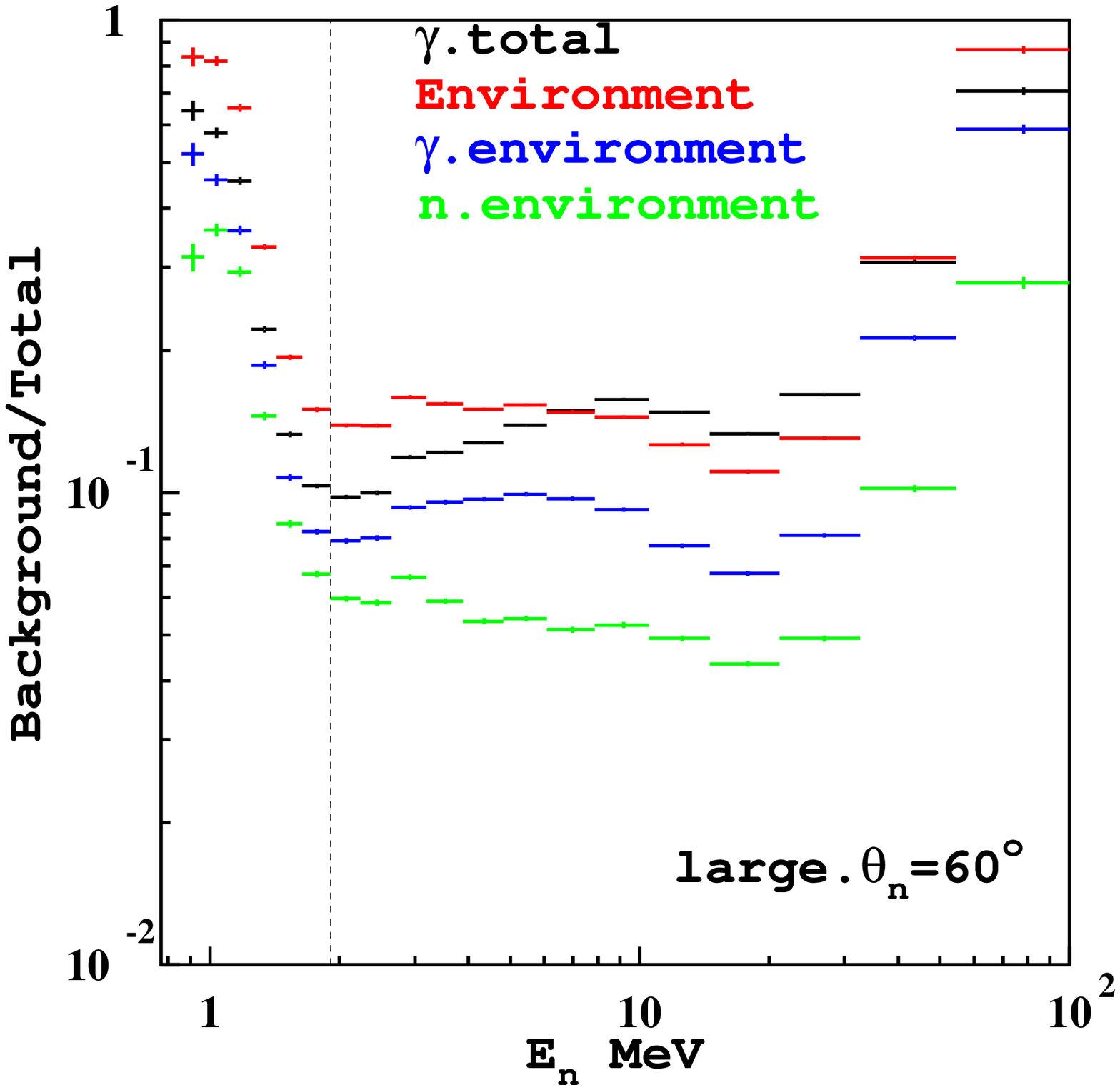}
\caption{\label{fig:bkg_r}Background contributions for far detectors.}
\end{minipage}
\end{figure}
The relative contributions of these two backgrounds are shown in Figs.~\ref{fig:bkg_l} and \ref{fig:bkg_r}.
Contributions of both backgrounds are of the order of 10\%, thanks to the choice of a light beamline
material and the empty floor of the experimental hall. All background contributions exhibit rapid rise at low and high
neutron energy endpoints. The reason is the fast drop of the measured neutron yield, due to the low efficiency
at small neutron energy and because of the phase space shrinkage at high neutron energy.
Near detectors saw considerably smaller backgrounds because of their smaller sensitive volume.
The environmental background in near detectors was dominated by neutrons rescattered from surrounding
materials, whereas in far detectors higher detection threshold suppressed them with respect to
environmental $\gamma$s.

\section{Results}\label{sec:res}
Fully differential neutron yields had been measured at 16 angular settings from 0 to 150 degrees
covering a wide neutron energy range from 0.5 MeV to the beam energy. Such a broad kinematic coverage
was necessary to obtain a complete information on the neutrons produced from $^9$Be target.
Indeed, the power of the ADS depend directly from the neutron yield and its
energy spectrum, demanding for a detailed and precise measurement. The beam time assigned by PAC~\cite{proposal}
was sufficient to acquire millions of events at each angular setting. Thus, except for
kinematic end points, the statistical uncertainties are negligible. The precision of the experiment
was limited only by the systematic uncertainties.

\begin{figure}[h]
\begin{minipage}{18pc}
\includegraphics[bb=20 180 580 640, width=18pc]{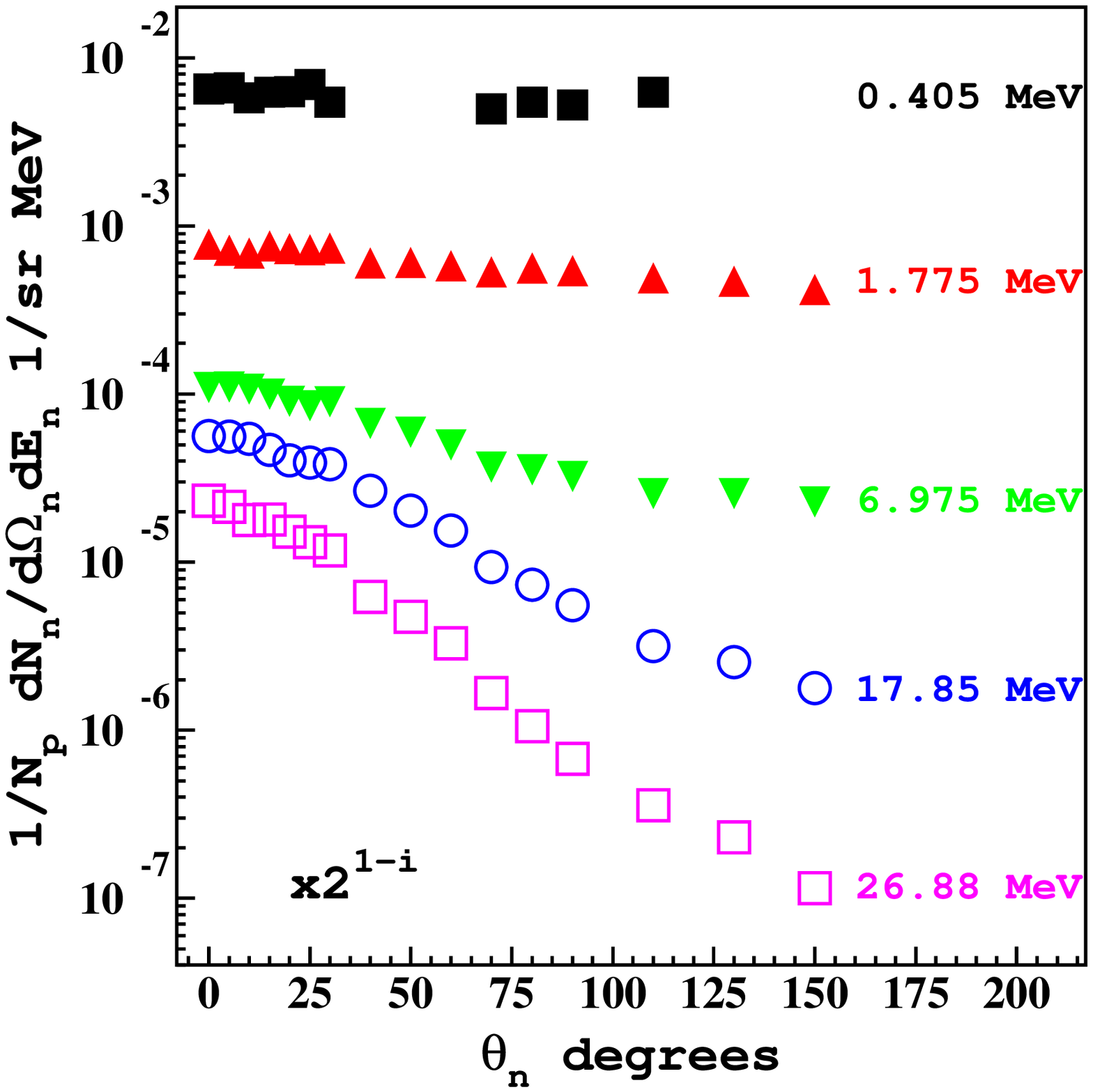}
\caption{\label{fig:diff_yield_l}Angular dependencies of measured differential yield,
rescaled in each successive energy bin by a factor of two for a better visibility.}
\end{minipage}\hspace{2pc}%
\begin{minipage}{18pc}
\includegraphics[bb=20 180 580 640, width=18pc]{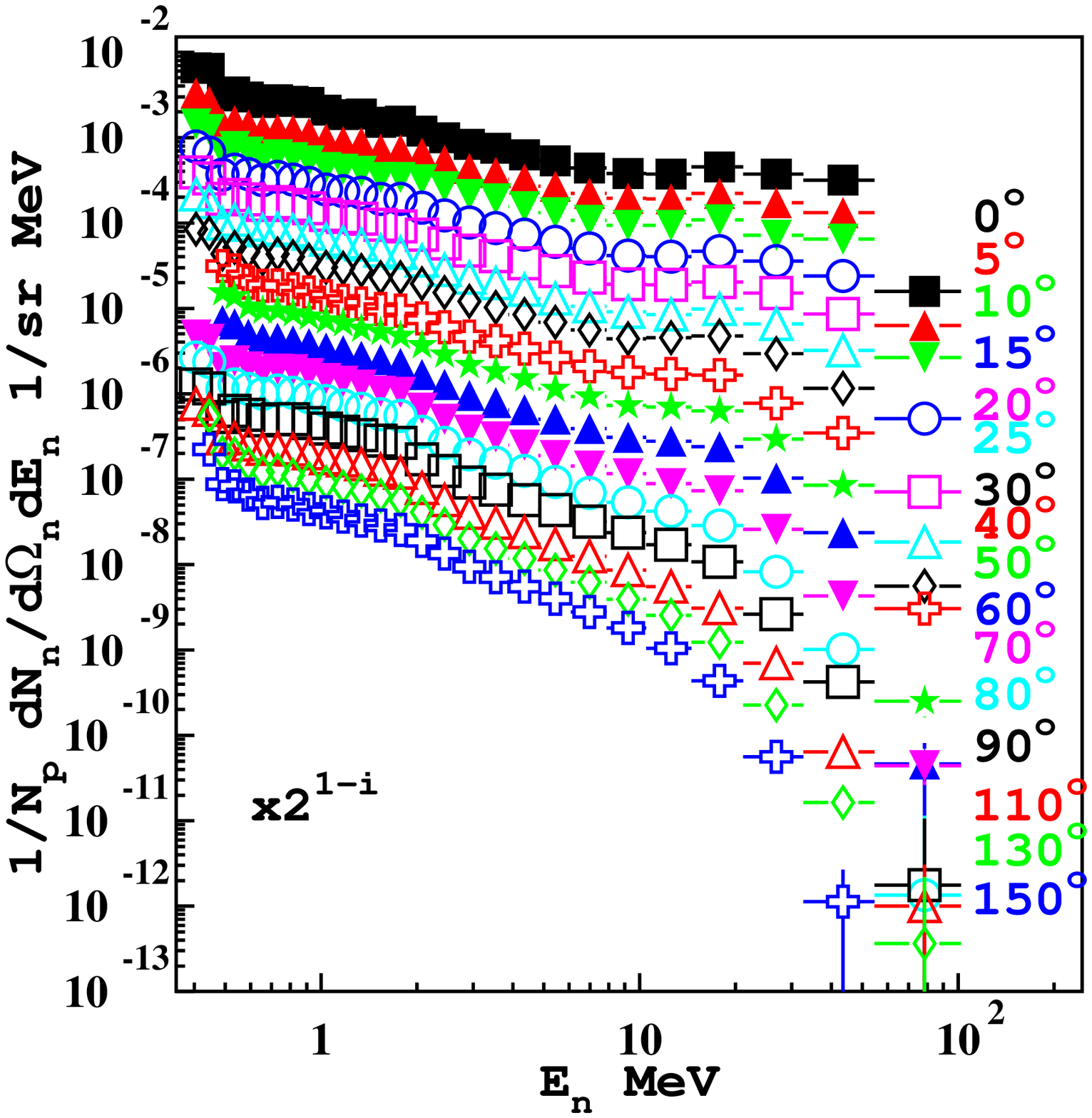}
\caption{\label{fig:diff_yield_r}Energy dependencies of measured differential yield,
rescaled in each successive angular bin by a factor of two for a better visibility.}
\end{minipage}
\end{figure}

Preliminary data on the differential angular and energy distributions are shown in Figs.~\ref{fig:diff_yield_l}
and \ref{fig:diff_yield_r}.

\begin{figure}[h]
\begin{minipage}{18pc}
\includegraphics[bb=20 180 580 640, width=18pc]{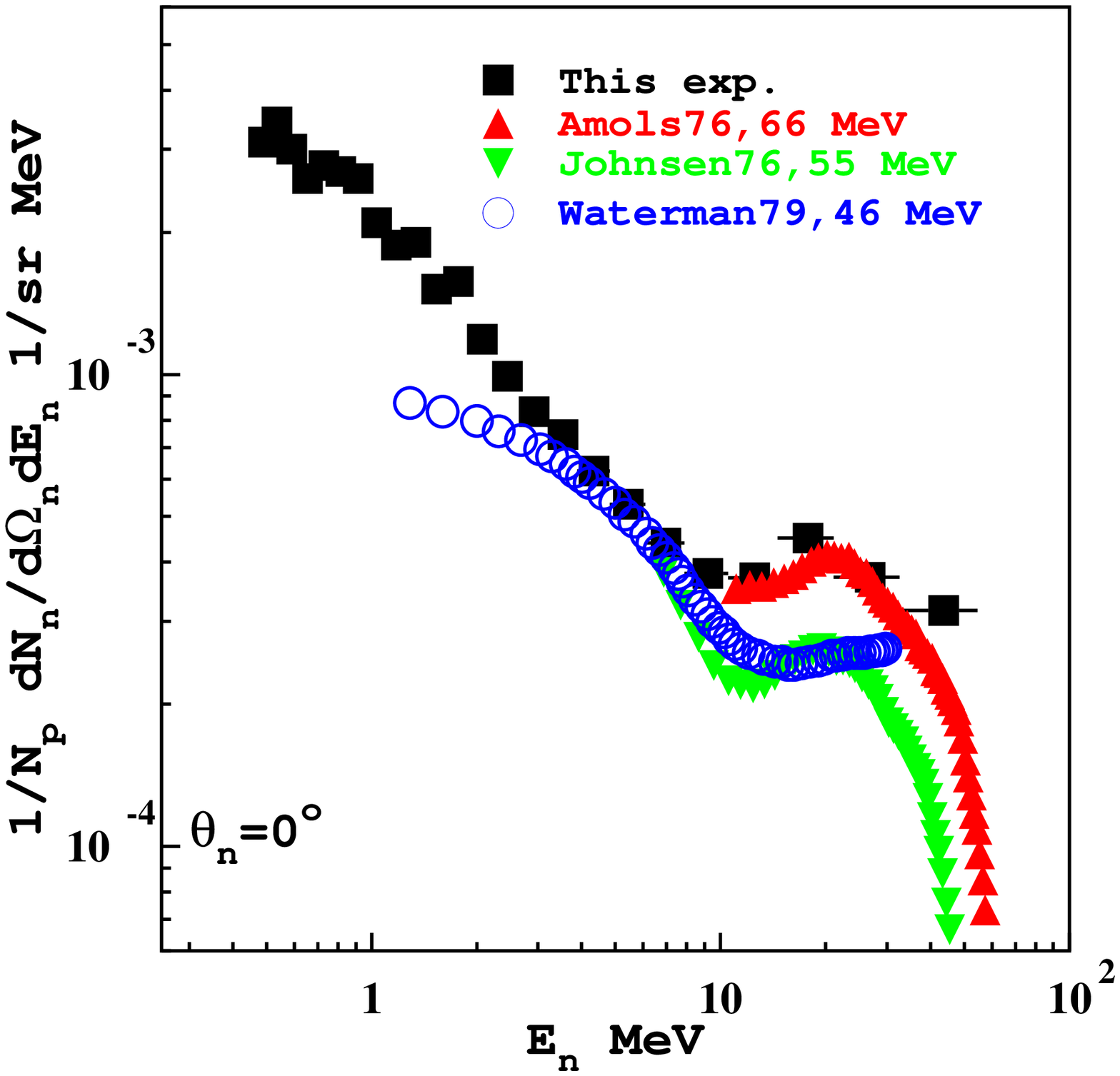}
\caption{\label{fig:world_l}Comparison to the world data at 0$^\circ$ from Refs.~\cite{Almos77,Johnsen76,Waterman79}.}
\end{minipage}\hspace{2pc}%
\begin{minipage}{18pc}
\includegraphics[bb=20 180 580 640, width=18pc]{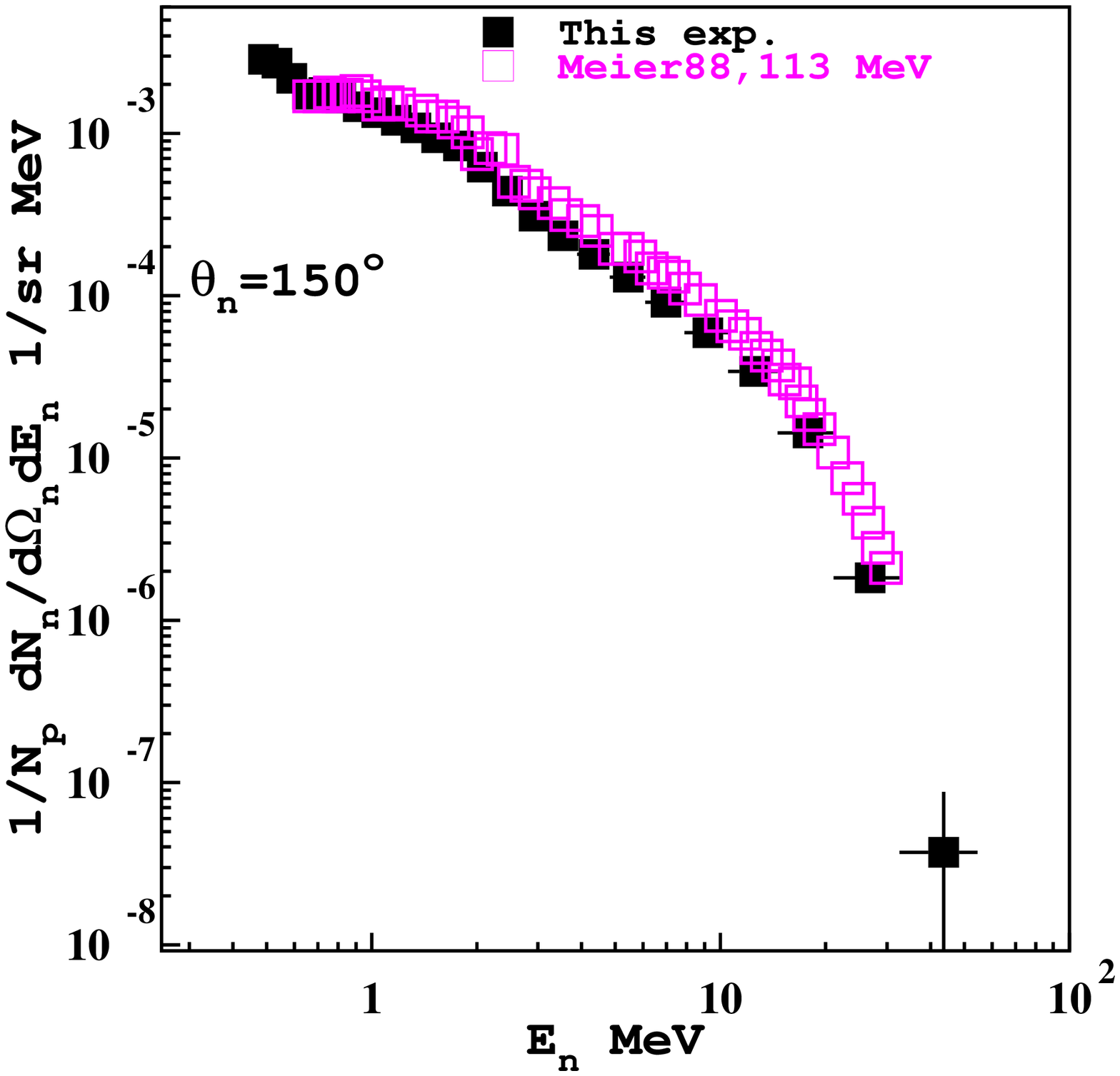}
\caption{\label{fig:world_r}Comparison to the world data at 150$^\circ$ from Ref.~\cite{Meier88}.}
\end{minipage}
\end{figure}

In Figs.~\ref{fig:world_l} and \ref{fig:world_r} these preliminary data were compared to the previous world data
from Refs.~\cite{Waterman79,Johnsen76,Almos77,Heintz77,Meier88,Madey77,Harrison80}.
The comparison to the data taken in Refs.~\cite{Waterman79,Johnsen76} with lower beam energy at 0$^\circ$
shows a significant disagreement where the neutron energy approaches the beam energy. This indeed was
expected as the lower beam energy reduces the available phase space. The disagreement demonstrates
also that the lower beam energy data cannot be extrapolated to higher beam energies simply
by an overall factor, but the kinematic limits have to be taken into account.
The comparison with Ref.~\cite{Waterman79} showed the disagreement also at neutron energies $<$3 MeV,
which could be attributed to the use of plastic scintillators in Ref.~\cite{Waterman79}.
Plastic scintillators do not exhibit PSD property and therefore $\gamma$/n-separation
cannot be performed. However, considering sign of the disagreement
the difference is more likely to be attributed to the detector efficiency calculations.
The data from Ref.~\cite{Waterman79} in the range from 3 to 8 MeV are in agreement with
our results. Our preliminary data are also in agreement with Refs.~\cite{Almos77}, which
shares almost the same beam energy, and with detailed data set from Ref.~\cite{Meier88} taken at 113 MeV.

\begin{figure}[h]
\begin{minipage}{18pc}
\includegraphics[bb=20 180 580 640, width=18pc]{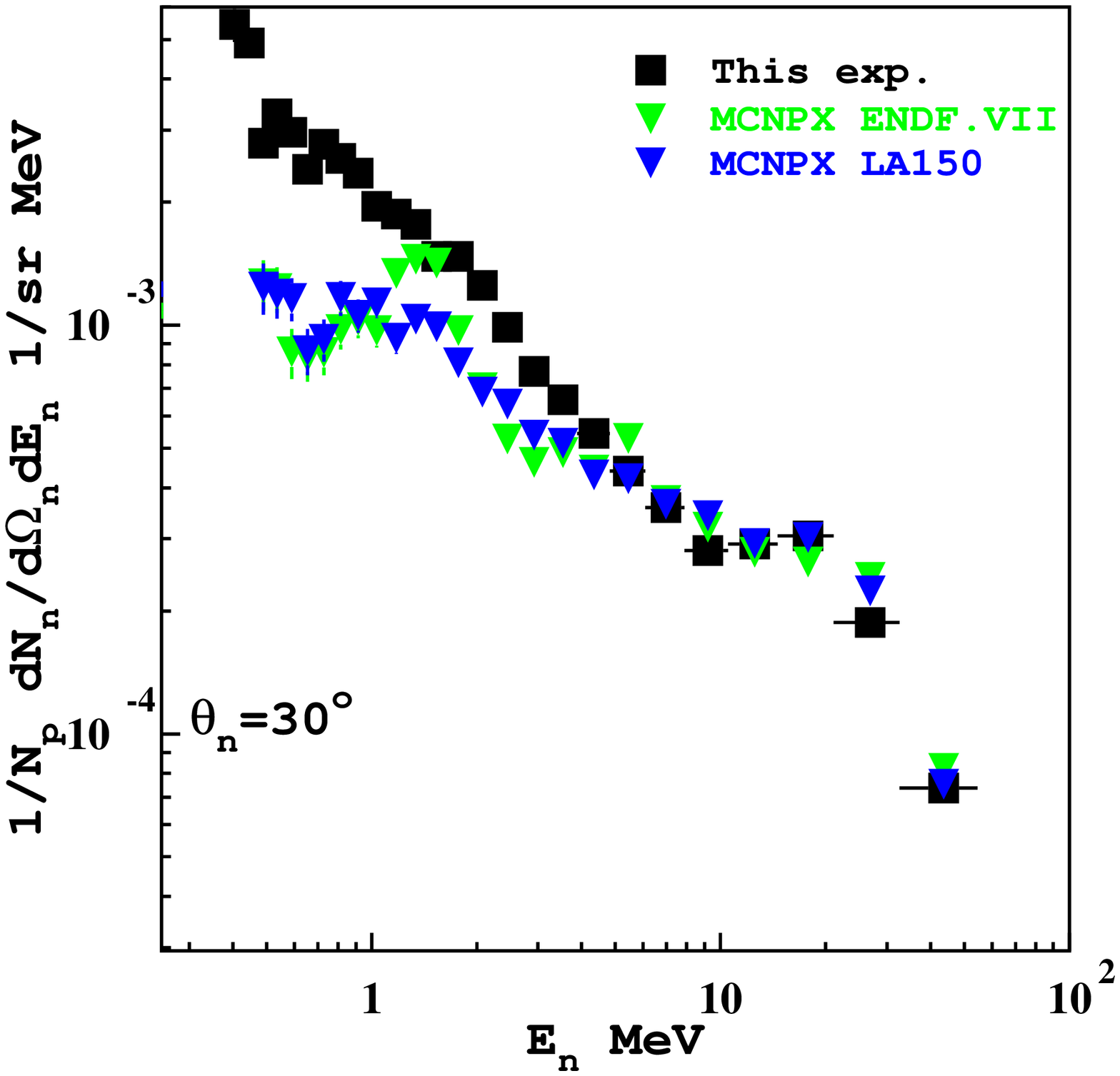}
\caption{\label{fig:mcnpx_l}Comparison to MCNPX at 30$^\circ$.}
\end{minipage}\hspace{2pc}%
\begin{minipage}{18pc}
\includegraphics[bb=20 180 580 640, width=18pc]{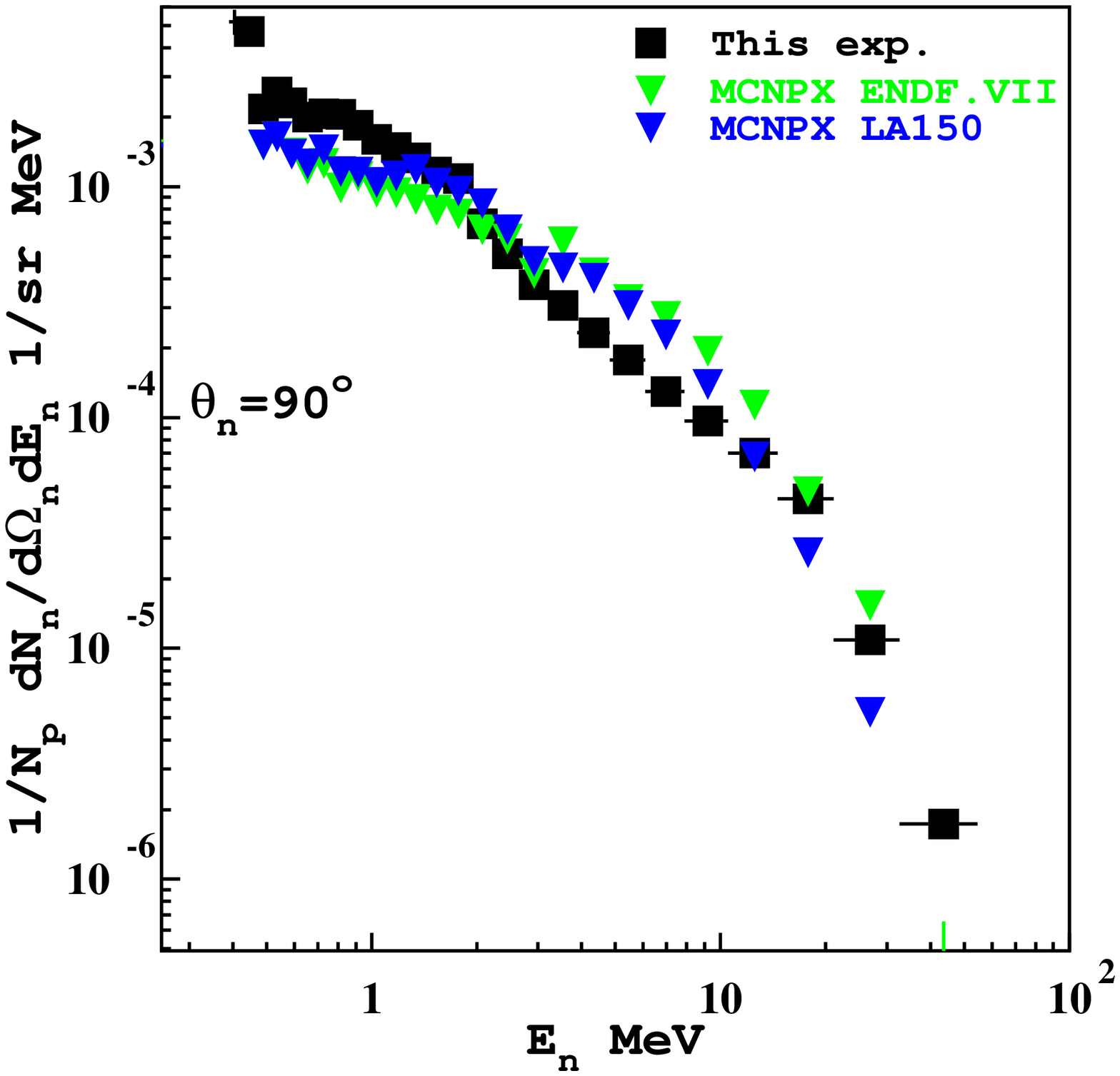}
\caption{\label{fig:mcnpx_r}Comparison to MCNPX at 90$^\circ$.}
\end{minipage}
\end{figure}

The data were also compared to MCNPX simulations as shown in Figs.~\ref{fig:mcnpx_l} and \ref{fig:mcnpx_r}.
The design of ADS was performed in MCNP. Hence it was mandatory to check the neutron yield simulations by this software package. The simulations were performed using two different physics libraries: ENDF VII and LA150.
The obtained comparison shows rather good agreement in some kinematic regions, like for small angles
and high neutron energy. At neutron energies below 1-2 MeV MCNPX yields lie below the data.

\begin{figure}[h]
\begin{minipage}{18pc}
\includegraphics[bb=20 180 580 640, width=18pc]{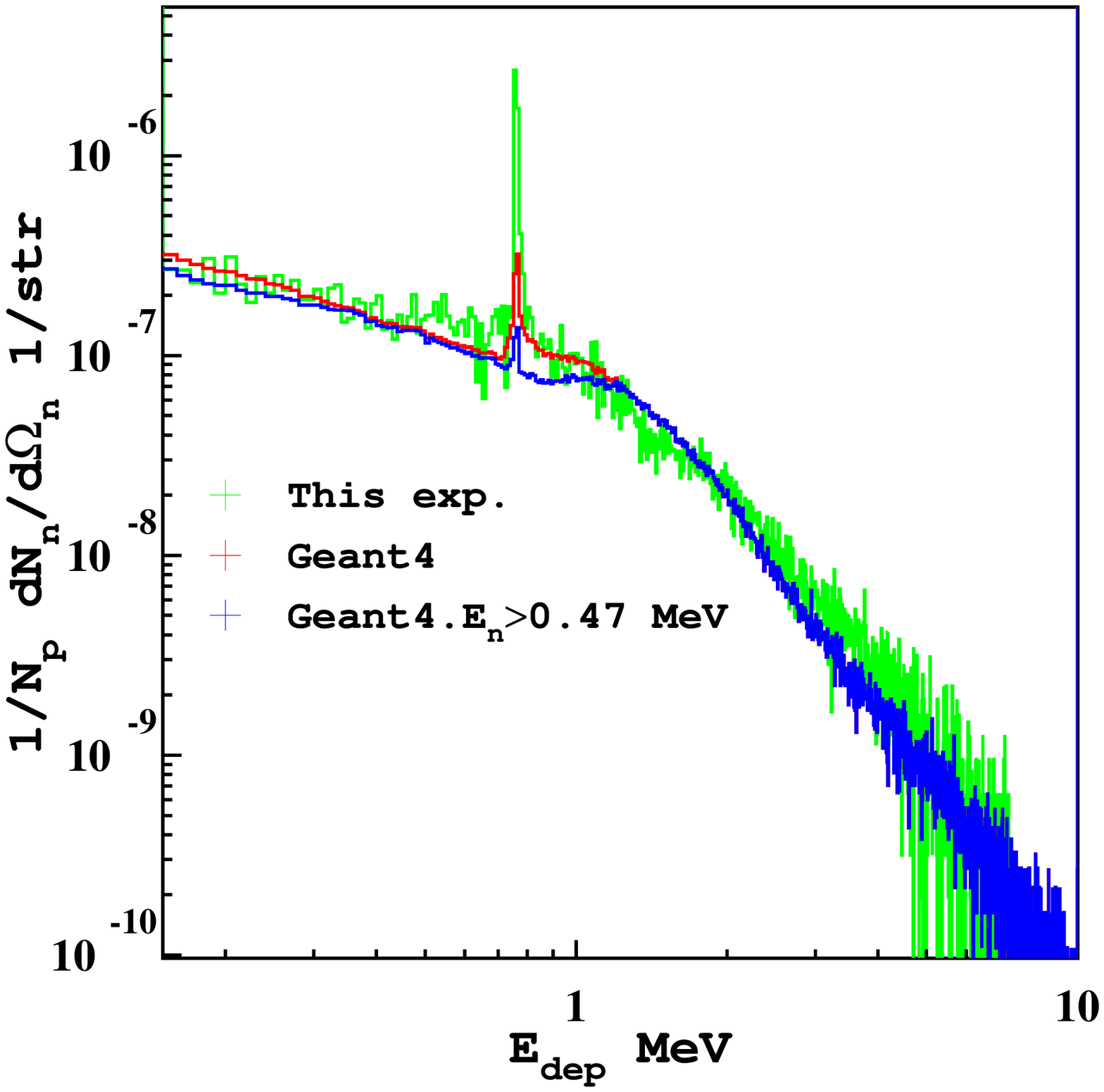}
\caption{\label{fig:he3_l}Deposited energy in $^3$He detector compared to Geant4 simulations.}
\end{minipage}\hspace{2pc}%
\begin{minipage}{18pc}
\includegraphics[bb=20 180 580 640, width=18pc]{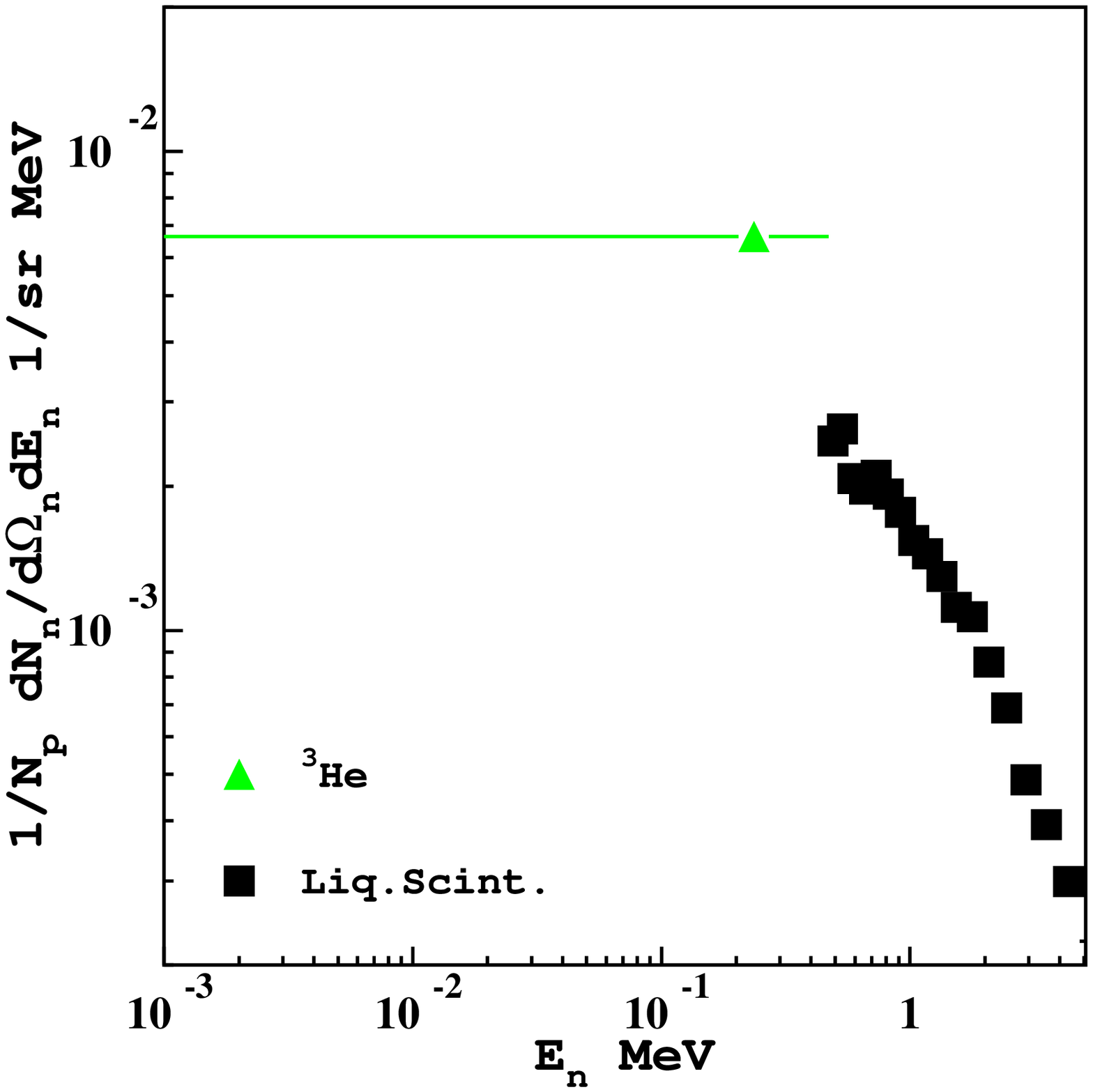}
\caption{\label{fig:he3_r}$^3$He detector yield in comparison with liquid scintillator. Uncertainties are statistical only.}
\end{minipage}
\end{figure}

Low energy neutrons $<$0.5 MeV could not be detected by liquid scintillators with a good efficiency and identification. These were measured
in special runs by $^3$He tube and Si detector covered with $^6$LiF converter. Preliminary data from $^3$He tube
are shown in Figs.~\ref{fig:he3_l} and \ref{fig:he3_r}.
The deposited energy spectrum shows a narrow peak due to the low-energy neutron conversion in proton
and tritium. Geant4 simulations indicate that interactions of high energy neutrons $>$0.5 MeV
in the $^3$He tube produce the background under the peak. These interactions occur mostly via
elastic scattering off $^3$He which does not convert whole neutron energy in deposited energy.
Moreover, higher energy products of neutron interaction inside $^3$He tube likely escape
the detector sensitive volume leading to partial loss of deposited energy.
This background of high energy neutrons was subtracted from the data and then the remaining yields
were corrected for $^3$He tube efficiency. The obtained neutron yield, integrated over the neutron energy
range from zero to 0.47 MeV, was compared to the liquid scintillator data in Fig.~\ref{fig:he3_r}.
In this preliminary analysis $^3$He data point lies above the nearest liquid scintillator data by a factor of two.
This point has to be further verified and compared to the data from Si detector because
the environmental background subtraction procedure introduces very large systematic uncertainty,
not discussed in this article.

\begin{figure}[h]
\includegraphics[bb=20 180 580 640, width=18pc]{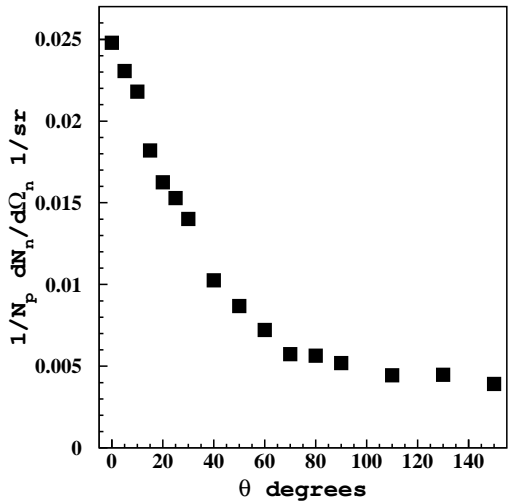}
\hspace{2pc}%
\begin{minipage}[b]{18pc}\caption{\label{fig:int_yield}Yield integrated over neutron energy.}
\end{minipage}
\end{figure}
Combining liquid scintillator data and $^3$He tube data the complete neutron energy coverage was achieved.
Thus the integrated neutron yield was calculated separately for each angular setting.
The obtained angular distribution of the integrated neutron yield is shown in Fig.~\ref{fig:int_yield}.
The integrated neutron yield falls rapidly with the angle due to shrinkage of the solid angle.
Integrating in turn the neutron yield in Fig.~\ref{fig:int_yield} also over the angle the total neutron
yield of $Y=0.1174\pm0.0002$ n/p was obtained. This value can be compared to the 
MCNPX simulations $Y=0.103$ n/p (ENDF VII) or $Y=0.096$ n/p (LA150) performed for the same beam energy,
and to the experimental data from Ref.~\cite{Tilquin05} $Y=0.110\pm0.007$ n/p at 65 MeV.
The experimental data from Ref.~\cite{Tilquin05} were measured by thermalization of produced
neutrons and their subsequent capture.

\section{Summary}\label{sec:summary}
Fully differential neutron yield produced by 62 MeV proton beam on a thick, fully absorbing $^9$Be target
was measured in almost complete energy and angular ranges.
The obtained data are in agreement with previous experiments
and show some deviation from MCNPX.
Estimates of systematic uncertainties are in progress,
but the expected values are of the order of 10\%.


\ack
Authors would like to acknowledge excellent support provided during the experiment
by the accelerator staff and technical services of Laboratori Nazionali del Sud.
This work was supported by the Istituto Nazionale di Fisica Nucleare INFN-E project.


\section*{References}
\bibliography{osipenko}

\end{document}